\documentclass[reprint,graphicx,
]{revtex4-1}

\usepackage{graphicx}

\usepackage{amsfonts,amssymb,amsmath}

\usepackage{color}

\begin{document}





\title{Persistent Ion Beam Induced Conductivity in Zinc Oxide Nanowires} 








\author{Andreas Johannes}

\email[]{andreas.johannes@uni-jena.de}

\author{Raphael Niepelt}

\author{Martin Gnauck}

\author{Carsten Ronning}




\affiliation{Institute of Solid State Physics, Friedrich-Schiller-University Jena, Max-Wien-Platz 1, 07743 Jena, Germany}





\date{\today}

\begin{abstract}

We report persistently increased conduction in ZnO nanowires irradiated by ion beam with various ion energies and species. This 
effect is shown to be related to the already known persistent photo conduction in ZnO and dubbed persistent ion beam induced conduction. Both effects show similar excitation efficiency, decay rates and chemical sensitivity. Persistent ion beam induced conduction will potentially allow countable (i.e. single dopant) implantation in ZnO nanostructures and other materials showing persistent photo conduction.

\end{abstract}


\maketitle 




The one dimensional geometry of nanowires leads to an immense surface to volume ratio and the path of any current through the wire is always close to the surface. Sensing applications \cite{Shen2009,Patolsky2004} benefit directly from this pronounced surface influence on electrical characteristics. Metal oxides in general and ZnO thin films in particular have already shown a good ultra-violet (UV) \cite{Emanetoglu2004} and gas-sensing \cite{Sberveglieri1995} ability due to the instability of oxygen at the surface \cite{Li2009,Prades2008,Bao2011}. As the surface influence is increased in nanowires, greatly enhanced sensitivity is expected from nanowire devices  \cite{Prades2008,Shen2009}. Although the conductivity of ZnO can easily be increased by three orders of magnitude by UV excitation, the current only returns to the relaxed value very slowly due to the known \underline{P}ersistant \underline{P}hoto \underline{C}onduction (PPC) effect \cite{Claflin2006,Polyakov2007,Prades2008,Zhou2009}. 

In this work, we report that an analogue \underline{P}ersistant \underline{I}on beam induced \underline{C}onduction (PIC) arises during ion irradiation. PIC may allow single ion detection in individual ZnO nanowires, as well as countable doping with single dopants into ZnO nanowires. The strongly localized energy deposition of a single ion impact can result in a change in the properties of the entire device, providing a localized probe to understand the underlying excitation mechanisms. 

The typical lifetime of charge carriers in ZnO nanostructures is in the order of 100\,ps \cite{Reparaz2010}. To contribute to a \emph{persistent} current increase, excited charge carriers have to populate distinct, stable states. According to Prades et al. \cite{Prades2008} these are surface-oxygen bound states. Further discussion on the origin and properties of PPC can be found in literature \cite{Claflin2006,Polyakov2007,Prades2008,Zhou2009,Shen2009,Bao2011,Lany2005,White2011}. For PIC discussed in this work, ions instead of UV photons generate electron hole pairs in the semiconductor. By qualitative comparison we show that PIC and PPC rely on the same mechanism.

\begin{figure}
	\includegraphics{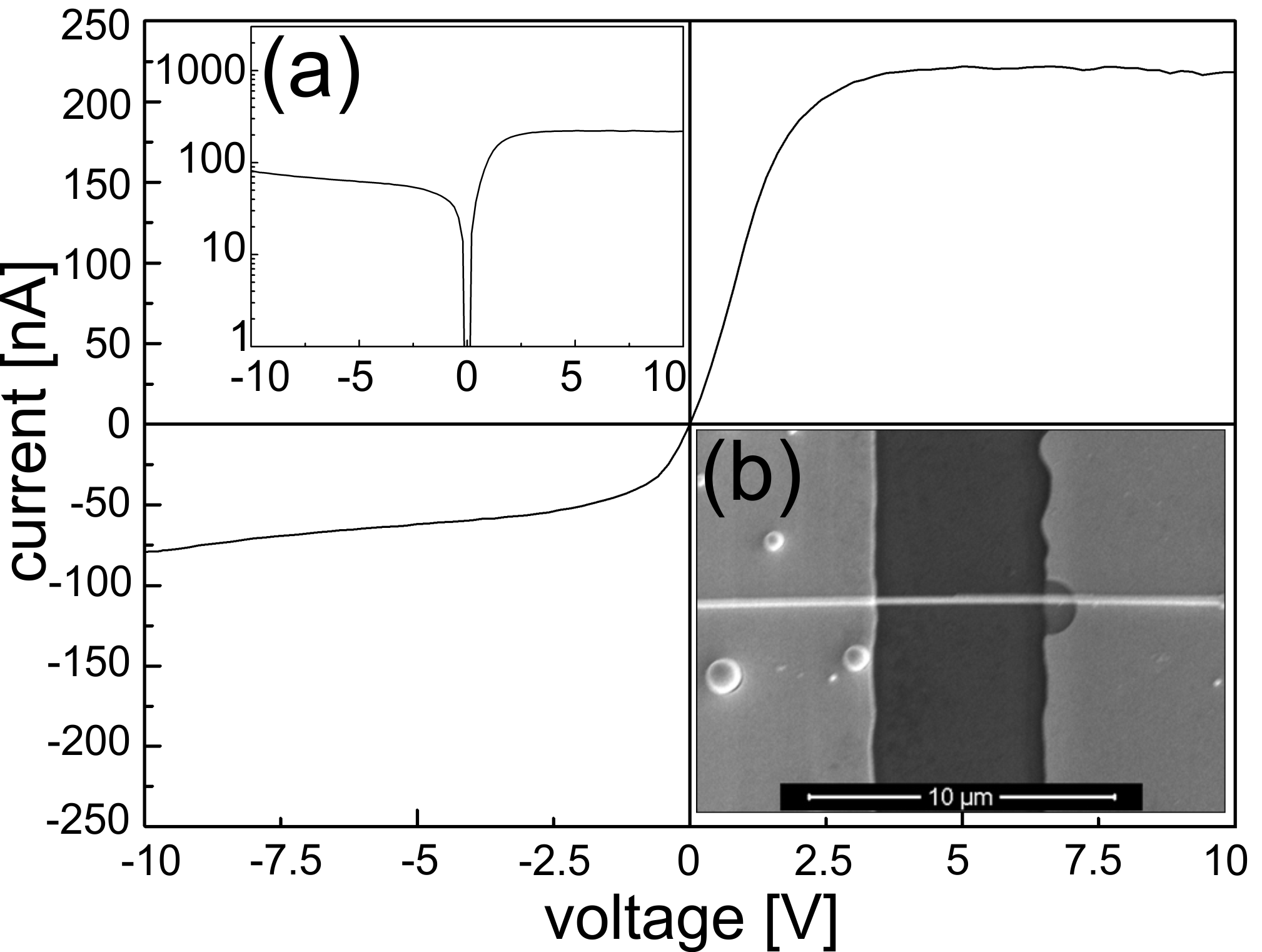}

	\caption{\label{pic1} \textit{I-V} response of a typical device investigated [in inset (a) as a logarithmic plot]. The device characteristics are dominated by the Schottky contacts formed between each end of the ZnO nanowire and the Ti/Au contact pads. Inset (b) shows an SEM image of a ZnO nanowire device investigated.}%
\end{figure}

The ZnO nanowires used for this work were grown via vapor-liquid-solid growth \cite{Wagner1964,Wacaser2009,Borchers2006} and have a typical diameter range of $150\,$ to $250$\,nm. They were imprinted onto the desired substrate by lightly pushing the substrate across the densly covered growth sample. Via photolithography and electron-beam evaporation, Ti/Au 50/50\,nm contact pads were defined onto the sparsely covered substrate. Some wires span the 5\,$\mu$m space between contact pads (see inset figure \ref{pic1}). Superfluous wires were easily cut with a focused ion beam to reliably manufacture single, contacted nanowires with this random-success method. The nanowire devices were then contacted and mounted onto a high vacuum flange fitted with electrical feed-throughs for in-situ characterization of the devices in the implantation chamber of a conventional 400\,keV multipurpose Implanter. All electrical characteristics were measured with a Keithley Source Measurement Unit (SMU model 237).

Figure \ref{pic1} shows the typical, strongly asymmetrical \textit{I-V} characteristics and a SEM image of such a single nanowire device, as manufactured.  The asymmetry, caused by high Schottky barriers forming between the intrinsic ZnO and the Ti/Au contacts, is often only exposed when measuring the characteristics across a sufficiently large voltage range, at least larger than the 3,4\,V corresponding to the band gap of ZnO. The current signals shown in further graphs were measured at a constant 1\,V bias. As reported in \cite{Bao2011} and elsewhere, our ZnO devices also show PPC and significant surface sensitivity, so that for comparable measurements the devices have to be kept in clean, dark ambient for at least 12\,h prior to experiments.

The devices were irradiated with a wide range of energies, ion-currents and total fluxes, while measuring their conductance at constant bias. Figure 2\,a) shows the current over time through a nanowire device, measured during two short 12\,s implantations of 5$\times 10^{10}\,$cm$^{-2}$ He$^+$ ions with 30\,keV at 0,6\,nAcm$^{-2}$ ion current. The remarkable increase in conduction from $\approx 0,5\,\mu$A to $\approx 2,5\,\mu$A during the first implantation at 1000\,s is reminiscent of many PPC experiments with ZnO using UV LEDs (see eg. figure 2\,b) or \cite{Bao2011}). A second identical implantation step at $3000\,$s leads to a similar increase from $\approx 2\,\mu$A to $\approx 4\,\mu$A. Further implantation would eventually saturate the current on a level varying from device to device. If the high vacuum is maintained, the current decays steadily, but extremely slowly, not reaching the original, relaxed value in any accessible time. At $6000$\,s the implantation chamber was vented, leading to much faster conductivity decay in air and dark. 

Figure 2\,b) shows the decay of PIC and PPC in different atmospheres. This graph is included to show that the exponential-like decay rate and chemical (O$_2$, H$_2$O) sensitivity of PIC is analogous to that of PPC. Similar experiments were conducted in noble gas environment, to exclude a pure pressure dependence (from loose contacts etc.). Apart from an overall increased conductivity, the \textit{I-V} characteristics retained their asymmetric shape (shown in figure 1) after implantation.

\begin{figure}
	\includegraphics{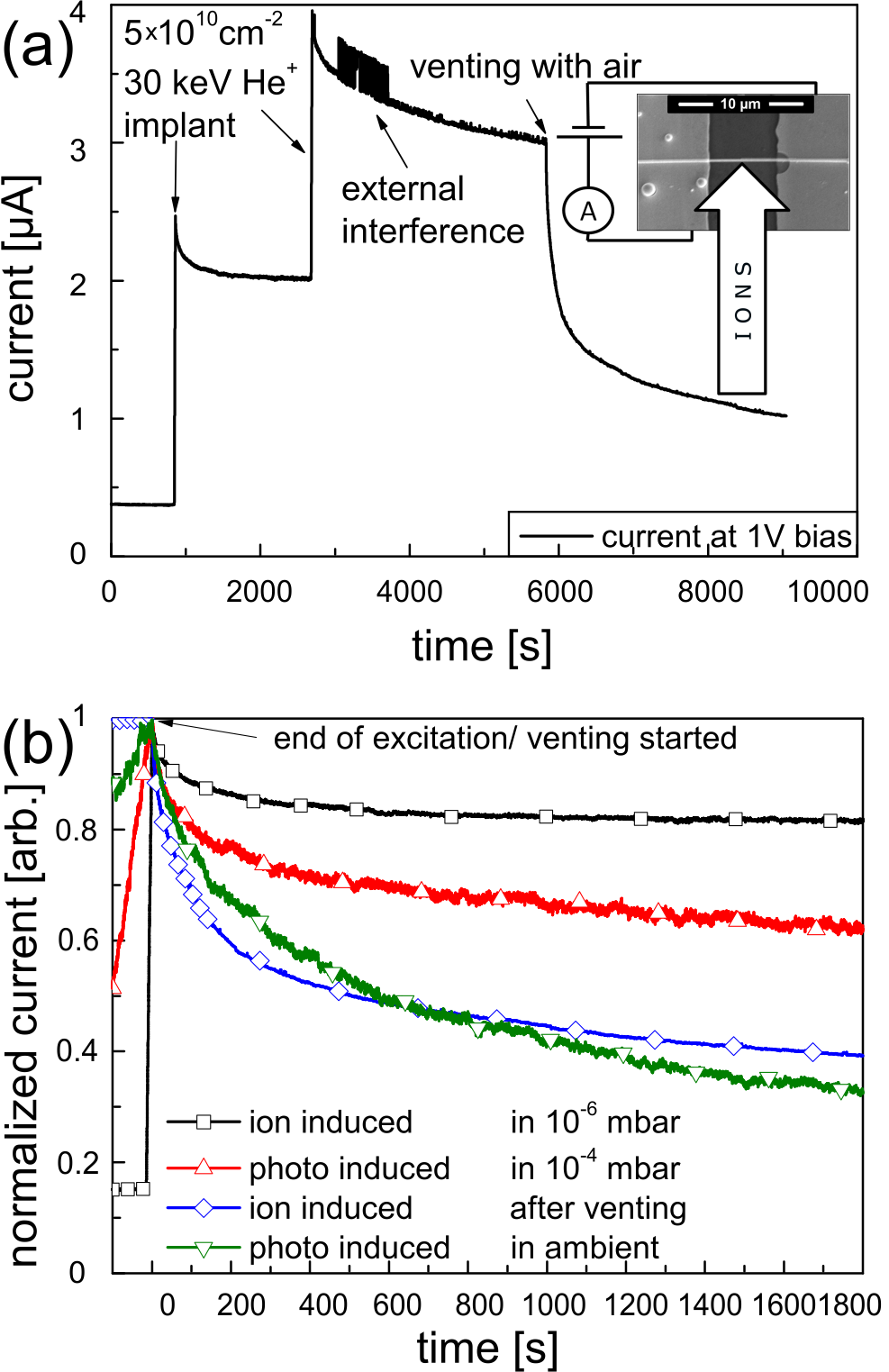}

	\caption{\label{pic2} Figure (a) is a plot of the current measured through a nanowire device at 1\,V bias during two short exposures to 5$\times10^{10}\,$cm$^{-2}\,30\,$keV He$^+$ ions at $200$\,nA, as shown in the inset. After 1,5\,h the implantation chamber was vented with air, causing a marked increase in the decay rate of the current. The stark increase and slow decay in conductivity of a ZnO nanowire induced by ion beam irradiation is reminiscent of the PPC effect. The disturbance marked "external influence" is attributed to the SMU changing detection ranges. Figure (b) illustrates the chemical sensitivity of the exponential-like decay times and their remarkable resemblance to PPC. The maximum current through two different devices at 1\,V is normalized to 1. The strong similarity of PPC and PIC behavior suggests the same underlying effect. 
}

\end{figure}

From the SEM image of the device (see figure \ref{pic1}) the area of the nanowire irradiated can be estimated to 200\,nm$\times 5\,\mu$m = $1\,\mu$m$^2$. The ion fluence of 5$\times 10^{10}\,$cm$^{-2}$ He$^+$ ions thus corresponds to roughly 500 ions per nanowire. The increase in conductivity of $2\,\mu$A is thus due to the impingement of roughly 500 ions, corresponding to an increase of roughly 4\,nA per ion. This is well within the resolution of the SMU and in the order of magnitude of the current conducted by relaxed single nanowire devices, so that the detection of a single event is plausible.

Helium was chosen as ion species for the first experiments, as doping and lattice damage can be excluded as cause for any electrical response. The impinging He$^+$ ions lose their charge and diffuse out of the lattice and virtually no defects are generated by the light He$^+$ ions at this energy \cite{Lorenz2005,Ronning2010,Borschel2011} due to strong dynamic annealing in ZnO.

Each impinging ion has 30\,keV of energy. The absolute maximum number of charge carriers which can be generated with this energy is 2$\times$30\,keV$ / 3,4\,$eV$ \approx$ 18 000 (2$\times$ion energy/band gap). Note that this is a gross overestimation of the charge carriers \cite{Klein1968}. We find, that an ion current of $0,6\,$nAcm$^{-2}$ thus generates a maximum charge carrier density of $\approx 3\times10^{8}\,$cm$^{-3}$, assuming a lifetime of $100\,$ps \cite{Reparaz2010}. With a mobility of 20\,cm$^{2}/$Vs \cite{Bao2011} and a bias of 1\,V across the nanowire length of $5\,\mu$m this amounts to an directly ion induced current of roughly $1\,$fA, far less than the observed current change. Although it was already clear, that the charge carriers generated by ion excitation do not have a sufficient life time to contribute \emph{directly} to \emph{persistent} conductivity changes, this estimation shows the exaggerated current excitation efficiency of PIC in ZnO. Similar results have been published for the PPC effect (see for eg. \cite{Bao2011}).

\begin{figure}
	\includegraphics{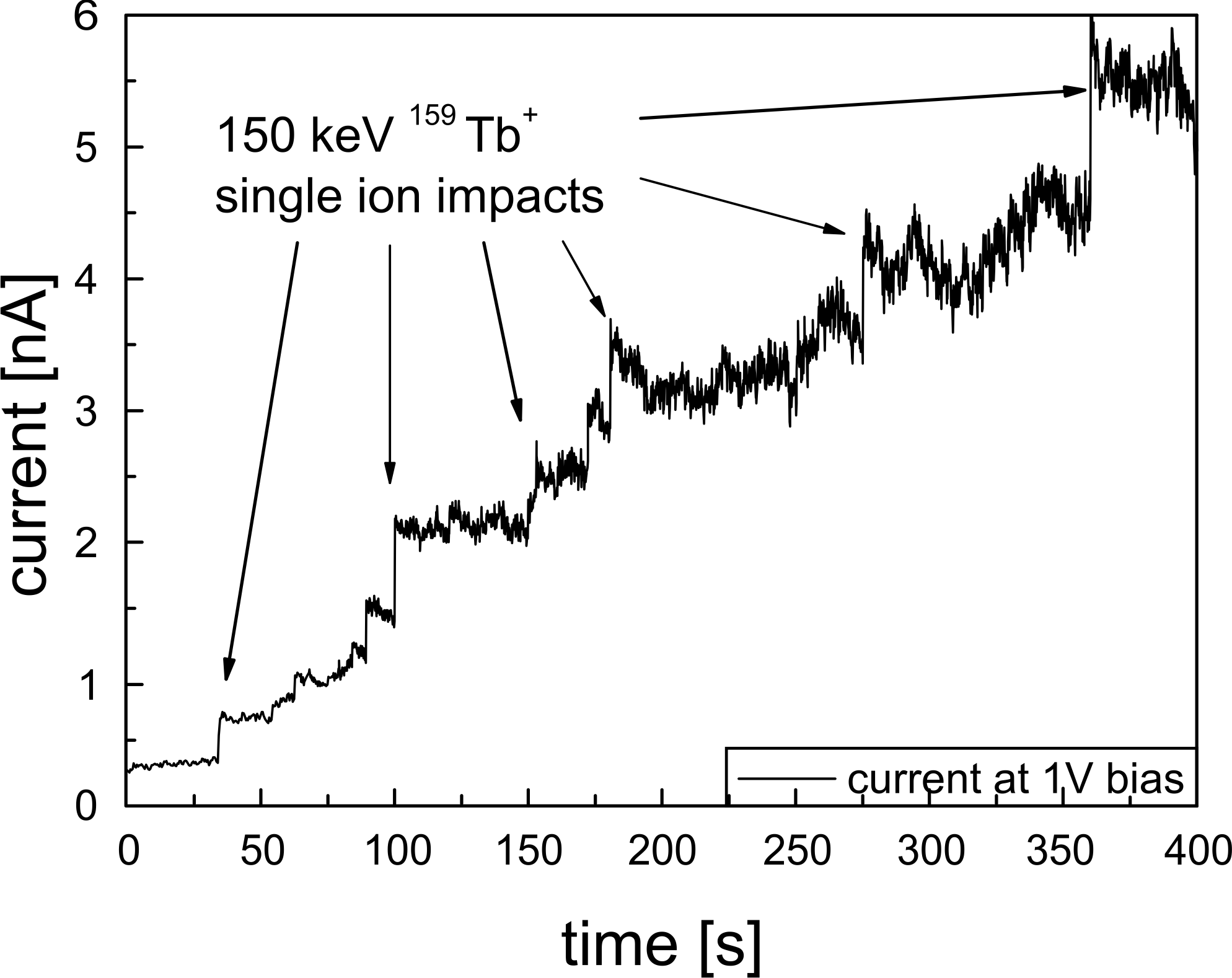}

	\caption{\label{pic3}Current through a nanowire device at 1\,V during strongly attenuated ion beam exposure (ion current $\approx5\,$pA/cm$^{-2}$). The single incidences of increasing conductivity indicate sufficient sensitivity of relaxed devices to detect single ions.}
\end{figure}

Figure \ref{pic3} shows the current through a ZnO nanowire device during implantation, this time at a minimal ion current of $\approx 5\,$pAcm$^{-2}$ terbium ions. This corresponds to an estimated one impingement per three seconds exposure. A fully relaxed device with low initial conductivity of $\approx 0,5\,$nA increases its conductivity to 6\,nA in distinct steps during $\approx 350\,$s of exposure. Terbium was used for this implantation to show that the PIC effect is not at all ion specific and because rare earths show interesting optical effects in ZnO \cite{Geburt2008}.

It was estimated that about 30-40 ions would impinge the exposed nanowire, however only 9 distinct events could be detected. This may be attributed to some ions merely glancing off the nanowire, not depositing sufficient energy for a marked conductivity increase. Also not the entire area of the nanowire may respond equally to excitation. As the devices' \textit{I-V} characteristics are dominated by Schottky barriers, any changes in these areas will have the greatest effects. It is plausible, that the area near the metal contacts, where the Schottky barriers are formed, responds more sensitively to excitation. Ions impinging the center of the nanowire device would then be largely without effect. The strongly localized excitation of fast ions thus shows the importance of the contact area for device characteristics. Other groups have also reported vastly increased UV sensitivity from Schottky type ZnO-metal contacts \cite{Zhou2009}. The prevailing model for PPC may have to be expanded to allow for the band bending induced at the ZnO-metal contacts, as the Schottky barrier adds an additional, uncontrolled complexity to the device properties. 

In conclusion, PIC has been shown to exist as an effect in ZnO closely related to PPC. Excitation efficiency, decay rates and chemical sensitivity of both effects match. The PIC effect is predicted to also occur in all other materials that exhibit PPC, as the energy brought into the material by ions can activate (meta-)stable states similarly to photons. PIC can thus help to expand the understanding of other materials' response to external excitation. PIC can potentially lead to countable implantation in ZnO and other materials showing PPC. ZnO nanowires provide a large set of parameters of which some are difficult to even control qualitatively. However, the pronounced sensitivity of ZnO nanowire devices shown and exploited in this work will surely lead to superior sensing applications.









%




%











%











%

\end{document}